\documentclass[twocolumn,twoside]{Jornadas}
\usepackage[utf8]{inputenc}
\usepackage{listings}
\usepackage{algorithm}
\usepackage{algorithmicx}
\usepackage{algcompatible}
\usepackage{adjustbox}
\usepackage{graphicx}
\usepackage{color}
\usepackage{caption}
\usepackage{comment}
\usepackage{amsmath}
\usepackage{amsfonts}
\usepackage{amssymb}
\usepackage[caption=false,font=footnotesize]{subfig}
\usepackage{placeins}
\usepackage{hyperref}
\usepackage{url}
\usepackage{bm}
\usepackage{cite}

\definecolor{gray97}{gray}{.97}
\definecolor{gray75}{gray}{.75}
\definecolor{gray45}{gray}{.45}

\def\BibTeX{{\rm B\kern-.05em{\sc i\kern-.025em b}\kern-.08em
    T\kern-.1667em\lower.7ex\hbox{E}\kern-.125emX}}

\graphicspath{{.}{./Figuras/}}

\begin{document}

\title{An Information Theory Approach to Physical Domain Discovery}

\author{%
     Daniel Shea%
     \thanks{Department of Materials Science and Engineering, University of Washington},
     Stephen Casey%
     \thanks{NASA Langley Research Center; {\tt Stephen.Casey@nasa.gov}}
}

\maketitle
\markboth{}{}
\pagestyle{empty} 
\thispagestyle{empty} 

\begin{abstract}
The project of physics discovery is often equivalent to finding the most concise description of a physical system.  The description with optimum predictive capability for a dataset generated by a physical system is one that minimizes both predictive error on the dataset and the complexity of the description.  The discovery of the governing physics of a system can therefore be viewed as a mathematical optimization problem.  We outline here a method to optimize the description of arbitrarily complex physical systems by minimizing the entropy of the description of the system.  The Recursive Domain Partitioning (RDP) procedure finds the optimum partitioning of each physical domain into subdomains, and the optimum predictive function within each subdomain.  Penalty functions are introduced to limit the complexity of the predictive function within each domain.  Examples are shown in 1D and 2D.  In 1D, the technique effectively discovers the elastic and plastic regions within a stress-strain curve generated by simulations of amorphous carbon material, while in 2D the technique discovers the free-flow region and the inertially-obstructed flow region in the simulation of fluid flow across a plate.
\end{abstract}


\section{Introduction}
\PARstart{T}{he} Kolmogorov complexity of a mathematical equation - the length of the shortest computer program that produces the equation as an output - is relatively small.  The dataset of observations of physical process may be arbitrarily large, but if the underlying equation describing the process is known, the dataset may be regenerated using only the amount of information entropy contained in this function.   The process of finding a symbolic equation to describe a physical system is equivalent to reducing the minimum amount of information needed to describe the system.  

Symbolic regression is an algorithmic approach to searching the space of mathematical expressions in order to represent complex datasets with low-complexity symbolic descriptions.  There has been significant progress in recent years in developing AI-driven techniques to learn symbolic representations of physical systems from datasets \cite{Dzeroski2007}.  Areas of interest for these techniques include conservation laws \cite{Liu_2021}, differential equation models from multivariate time series \cite{Dzeroski_1995,BRADLEY2001139,105555,Langley_2015}, physics simulators \cite{watters2017visual}, an intuitive physics engine \cite{lake_ullman_tenenbaum_gershman_2017}, an automated adaptive inference agent \cite{daniels_2015}, and physical scene understanding \cite{Yildirim_2018}.  For arbitrarily large search spaces, genetic algorithms are often employed to manage the intractable combinatorial complexity encountered by brute-force approaches \cite{1010078,Pal_1996}.  The commercial Eureqa software \cite{Dubcakova_2011} uses the algorithm described in \cite{Schmidt_2009}.

Divide-and-conquer approaches can be applied with great effectiveness when the unknown function is a combination of known functions \cite{Wu_2019}.  Likewise, division of a physical domain into subdomains, or a dataset into subsets, is a powerful technique in identifying regimes where different physical processes are dominant.  These types of scenarios occur frequently in physics problems, such as the various flow regimes produced by the Navier-Stokes equations \cite{Batchelor_1967} or the elastic and plastic regimes in tensile deformation \cite{Callister_2004}.  Previous studies \cite{Wu_2019} propose developing physical theories and finding the domains in which these theories are applicable.  We take the reverse approach and identify the domains with the most succinct descriptions.  The partitions created by the process of minimizing subdomain description complexity are often highly effective at separating regimes dominated by different physical phenomena.

\begin{figure}
\centering
  \includegraphics[width=0.45\textwidth]{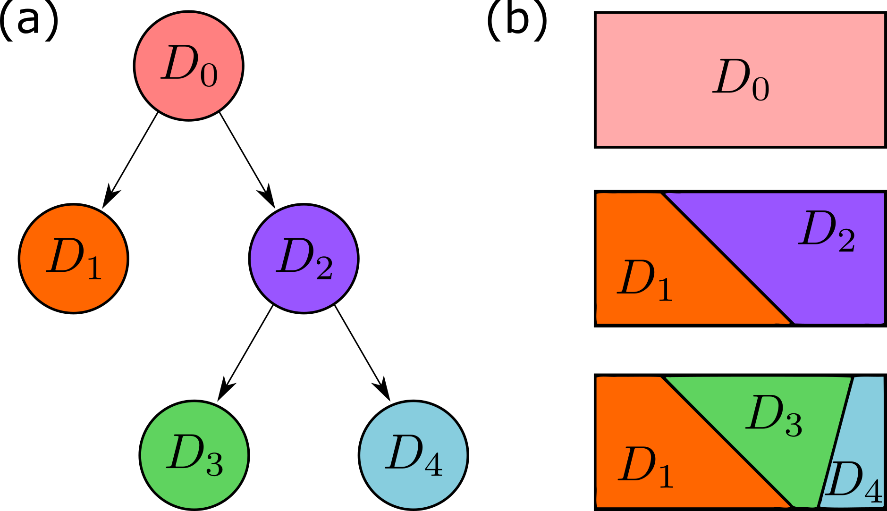}
  \caption{Subdomains are discovered via binary space partitioning.  A recursive search over previously-identified subdomains of the top-level domain $D_0$ is used to identify subdomains that minimize reconstruction error of the data and simultaneously minimize model complexity.}
  \label{fig:subvol}
\end{figure}

The process of dividing an N-dimensional space via an (N-1)-dimensional hyperplane, known as Binary Space Partitioning, has a long history of use in computer science \cite{Schumacker_1969}.  Finding the position of this cutting plane that minimizes the error and complexity metric of the system description can be formulated as an optimization problem.  While a brute-force method is employed here to find the location of the cutting plane and fully characterize the error function, in practice this problem has also proven amenable to optimization techniques such as simulated annealing, Monte Carlo methods, and Bayesian sampling.

Optimizing the accuracy of the model fit while minimizing the complexity of the description of each domain provides a formal information-theoretic implementation of Occam's Razor \cite{Jaynes_1957}.  When noise is present in the dataset, it is important to avoid overfitting as much as possible, thereby increasing the entropy of the predictive model.  Intuitively, the maximum-entropy description of a dataset makes the fewest assumptions about the data distribution and best represents the state of knowledge about the system \cite{Jaynes_1957}.  The binary partitioning procedure avoids the problem of overfitting and preserves close to the maximum entropy present in the noise.

Different complexity metrics may be used for different function and model types.  A wide range of function types may be used to describe a domain, such as symbolic equations, neural networks, or computer algorithms.  Examples of complexity metrics include Komolgorov complexity, the Vapnik–Chervonenkis dimension of a neural network \cite{Vapnik_2015}, the number of bits required to store an item in computer memory \cite{udrescu2020ai}, the number of terms in a symbolic equation, or any other metric best-suited to the problem.

\section{Methods}

\textbf{Problem and Notation:}  We define a discrete dataset $\bm{y_0}=y(\bm{x_0})$  for $\bm{x_0} \in \mathbb{R}^\textit{N}$.  The domain $D_0$ is defined to be the domain of $\bm{x_0}$.  In order to identify the optimal division of the original domain $D_0$ into a set of subdomains $D_i$ a binary space partitioning algorithm is used where a linear separation of a domain into two subdomains is performed at each step.  

A subdomain $D_i$ of $D_0$ contains a set of location points $\bm{x_i}$ where a dataset $\bm{y_i}=y(\bm{x_i}\mathnormal{})$ is defined. An approximation function or model $f_i(\bm{x_i}\mathnormal{})$ is used to fit the points in $\bm{y_i}$ as closely as possible.  In practice, $f_i$ may be an equation, algorithm, neural network, or any other function. An error metric $E(f_i,\bm{x_i},\bm{y_i}\mathnormal{})$, such as cumulative squared error, is used to measure how closely $f_i$ approximates $\bm{y_i}$.

The optimal approximation function $f_i^{(\mathrm{opt})}(\bm{x_i}\mathnormal{})$ will match the data $y_i$  as closely as possible, minimizing $E(f_i,\bm{x_i},\bm{y_i}\mathnormal{})$. However, in order to maximize predictive capability and avoid overfitting, the approximation function $f_i$ should approximate the data while minimizing effects of noise present in $y_i$, in accord with the principle of maximum entropy.  This principle states the probability distribution which best represents the current state of knowledge about a system is the one with largest entropy.  Overfitting is less likely to occur if the optimal model $f_i^{(\mathrm{opt})}$ has a low level of complexity, denoted by $C(f_i)$ where $C$ is a metric of the complexity of $f$.  $C(\cdot)$  measures the complexity of the function $f_i$ based on a relevant metric, such as the information entropy contained in the instructions for $f_i$, Kolmogorov complexity, Vapnik–Chervonenkis dimension of a neural network, number of terms in the model, or operations in a symbolic equation.  In some cases, the minimally complex representation is shown to be optimal in preserving as much predictive information as possible \cite{Tegmark_2019}.

A penalty function p is introduced to limit $C(f_i)$ by acting on the error $E(f_i,\bm{x_i},\bm{y_i}\mathnormal{})$. The effective error to be minimized is therefore\begin{equation}
E_{\mathrm{eff}}(f_i)=p(C(f_i)) \cdot E(f_i,\bm{x_i},\bm{y_i}\mathnormal{}).
\end{equation} The function p may be treated as a hyperparameter and adjusted for individual datasets.  The optimal function $f_i^{\mathrm{(opt)}}$ for a domain $D_i$ is given by \begin{equation}
f_i^{\mathrm{(opt)}}(\bm{x_i}) = \underset{f_i}{\mathrm{argmin}}\:E_{\mathrm{eff}}(f_i(\bm{x_i})).
\end{equation} For every subdomain $D_i$ there exists at least one optimized approximation function $f_i^{\mathrm{(opt)}}$.  The effective error produced by $f_i^{\mathrm{(opt)}}$ is defined as $E_i$.  Therefore each subdomain $D_i$ also has an associated optimized error $E_i$. 

The first step toward optimizing the description of the data contained in the top-level domain $D_0$ is to find the optimized approximation function $f_0^{\mathrm{(opt)}}$ and the associated error $E_0$.  Next, the optimal boundary hyperplane is identified which partitions $D_0$ into two subdomains $D_1$ and $D_2$ such that the cumulative error $E_1+E_2$ is minimized.  The boundary is a hyperplane of dimension $(N-1)$ for $D_0 \in \mathbb{R}^\textit{N}$.  Regardless of the location of this hyperplane,\begin{equation}
E_1 + E_2 \leq E_0.
\end{equation}This is because the models fit to the partitioned subdomains will always be as good or better than the model fit to the unpartitioned domain.  This type of max-cut problem  draws comparisons to other well-known problems in mathematics and computer science.

Once the optimal boundary hyperplane is identified, a pair of corresponding losses $E_1$ and $E_2$ can be computed. The final step is to determine if the identified boundary produces enough improvement to warrant a partition.  For this task a hyperparameter $q$ is introduced as a criterion to quantify the necessary improvement.  For a new boundary to be drawn the condition \begin{equation}
E_1 + E_2 \leq (1-q)E_0
\end{equation}must be satisfied.  In this work $q$ is set to 0.10 (or 10\%) unless otherwise specified. If a valid boundary hyperplane is identified, the partitioning process is repeated on the identified subdomains (in this case $D_1$ and $D_2$) until all subdomains are found.  This algorithm forms a binary partition tree as shown in Figure 1. 


The output of the RDP algorithm is a set of subdomains and models where the optimal model describing each subdomain is minimally complex with simultaneous maximal descriptive capability within its subdomain.  In Figure 1, the input data belongs to the domain $D_0$ and the algorithm identifies a set of subdomains $\{D_1,D_3,D_4\}$ that minimize reconstruction error of the data while simultaneously minimizing the complexity of functions used to fit the subdomains.
\section{Results}

\subsection{Implementation:  1D Example}
The simplest case for applying the approach described in the Methods section is a 1-dimensional dataset which contains two data domains. Consider the system \begin{equation}
y(x) = \left\{\begin{matrix}
50x^2-100x+250 & \:\:\:\:\; 0 \leq x < 10 \\
-50x-500 & \:\:\: 10 \leq x < 20
\end{matrix}\right.
\end{equation}Suppose there exists a dataset $\bm{y_0}=\bar{y}(\bm{x_0})+\epsilon(\bm{x_0})$ in the domain $D_0$ where the vector  $\bm{x_0}=[0, 0.01, 0.02, \cdots, 19.98, 19.99, 20.00]$ and $\epsilon(\bm{x_0})$ is white noise added to the data.  The noise is added by specifying a signal-to-noise (SNR) ratio with \begin{equation}
\text{SNR} = 10 \log_{10}\left ( \frac{\left \|\bar{y} (\bm{x_0})  \right \|_2^2}{\left \| \tilde{y} (\bm{x_0}) - \bar{y} (\bm{x_0}) \right \|_2^2} \right )
\end{equation} where $\bar{y} (\bm{x_0})$ represents the clean data and $\tilde{y} (\bm{x_0})$ represents the noisy data.  The noisy data $\bm{y_0}$ is shown in Figure 2 with $\mathrm{SNR} = 10$.

\begin{figure}
\centering
  \includegraphics[width=0.45\textwidth]{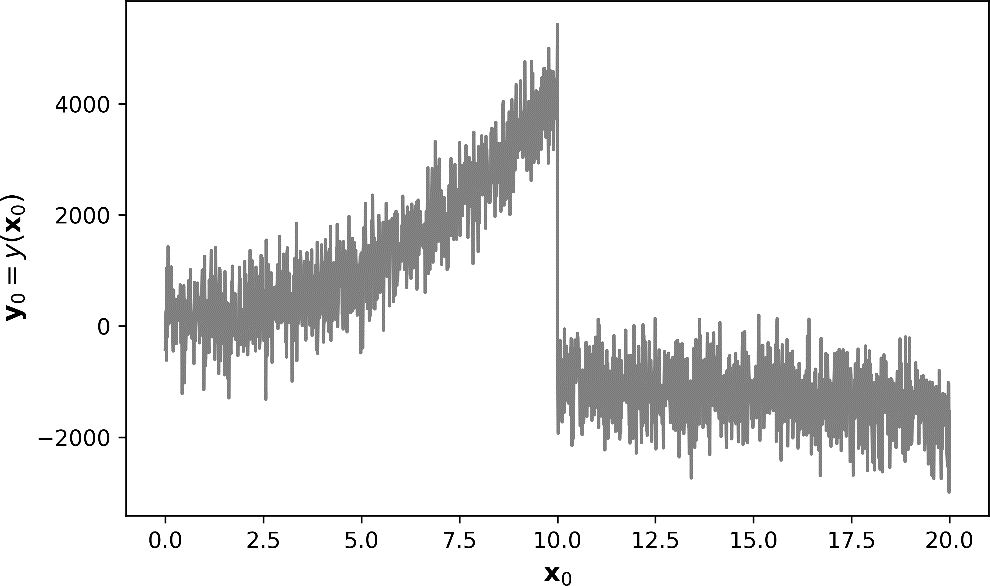}
  \caption{Example 1D dataset with two domains.  The system has a clear shift in its behavior at $x=10$.}
  \label{fig:subvol}
\end{figure}

Although a single model may be able to accurately capture the trends in the data, it will likely involve a model with a high degree of complexity that approximates the sharp change in the data values at $x=10$.  If the algorithm correctly identifies the boundary at $x=10$ where the true model changes, an accurate and parsimonious model may be learned on each subdomain $D_1 = [0,10)$ and $D_2 = [10,20]$.

A power series model is employed for the boundary and domain discovery algorithm, taking the form 
\begin{equation} f(x,K) = \sum_{k=1}^{K}a_kx^k \end{equation} where $a_k$ is the coefficient for the $k^{\mathrm{th}}$ term.  A larger value of K corresponds to greater model complexity.

The algorithm performs a brute-force search by evaluating a possible boundary at every point in $x_0$.  For each possible boundary placement, a set of models with varying complexity corresponding to the maximum exponent in the power series is fit to the data in each subdomain.  This example considers a set of functions \begin{equation} 
F = \left \{f(x,K) : K \in \mathbb{N}; \mathnormal{}K \leq \text{2} \right \}.
\end{equation}$K$ is set to be less than or equal to 2, corresponding to constant, linear, and quadratic polynomial models. The penalty function \begin{equation}
p\left ( C \left ( f(x,K) \right ) \right ) = 1 - 0.15(2-K)
\end{equation}is applied to the reconstruction loss $E\left (f_i, \bm{x_i},\bm{y_i} \right )$ for each of the proposed models in F, such that the error produced by a constant $f(x)=a_0$ model is multiplied by 0.70, the error produced by a linear model is multiplied by 0.85, and the error produced by a quadratic model is unchanged.  The reconstruction loss $E(\cdot)$ is the cumulative squared error for a model $f_i$ \begin{equation}
E\left ( f_i, \bm{x_i}, \bm{y_i} \right ) = \left \| y(\bm{x_i}) - f_i(\bm{x_i}) \right \|_2^2.
\end{equation} For each prospective boundary in the original domain $D_0$, the model $f_i$ which minimizes \begin{equation}
 E_{\mathrm{eff}}(f_i)=p(C(f(x,K))) \cdot E(f_i,\bm{x_i},\bm{y_i}\mathnormal{})  
\end{equation} is identified as the optimal model for the domain $D_i$.

The algorithm proceeds by placing a boundary, determining the optimal model for each of the subdomains defined by the boundary, computing the cumulative loss for each model, and summing together the cumulative losses $E_1 + E_2$.  In this example, the cumulative losses $E_1$ and $E_2$ are the effective losses for the domains $D_1$ and $D_2$ with data $y(\bm{x_1})$ and $y(\bm{x_2})$ respectively.  This procedure is repeated for each possible unique boundary that separates the data into two domains via a hyperplane of dimension $(N-1)$.  In the case of a 1D dataset, this hyperplane is a point, so each value in $\bm{x_0}$ is evaluated as a boundary.  Figure 3 presents the result of applying this procedure to three different prospective hyperplane boundary separation points in $D_0$. 

\begin{figure}
\centering
  \includegraphics[width=0.45\textwidth]{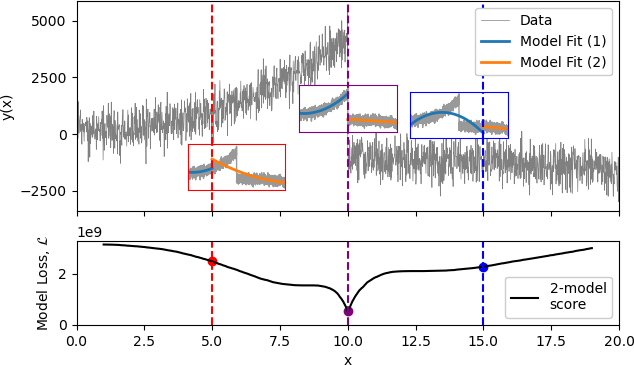}
  \caption{Identifying the boundary where the cumulative reconstruction loss is at a minimum (purple line).  The original data is shown in gray on the top part of the figure.  Three evaluation points and their corresponding reconstruction losses are shown in red, purple, and blue.}
  \label{fig:subvol}
\end{figure}

The center point $x=10$ is the best model separation point based on total cumulative model loss as shown in the bottom panel of Figure 3.  However, before we declare this hyperplane position as a boundary that separates two data subdomains, we must perform a comparison of the cumulative two-model loss score to the single-model loss score provided by fitting a model to the original dataset $\bm{y_0}$ in $D_0$.  As shown in equation 4, the hyperparameter q determines whether a boundary produces a sufficient improvement.  If $E_1+E_2 \leq (1-q)E_0$ is satisfied, the boundary is implemented and the data is divided into two subdomains.  For $q=0.10$ the boundary at $x=10$ satisfies this condition.

\subsection{Identifying Multiple Domains in Data}

The previous example demonstrates the RDP method on a 1D dataset with a single boundary.  In the case of multiple boundaries, the procedure finds multiple domains via a recursive search that iteratively places optimal boundary hyperplanes.  Consider a system with three distinct domains specified by the piecewise function\begin{equation}
y(x) = \left\{\begin{matrix}
\:\:\; 50x^2-100x+250, & \:\:\; 0 \leq x < 10 \\
-30x^2+150x-100, & \: 10 \leq x < 20 \\
-50x-500, & \: 20 \leq x \leq 30
\end{matrix}\right.
\end{equation} Suppose we have a dataset $\bm{y_0}=\bar{y}(\bm{x_0})+\epsilon(\bm{x_0})$ in the domain $D_0$ where the vector  $\bm{x_0} = [0, 0.01, 0.02, \cdots, 29.98, 29.99, 30.00]$ and $\epsilon(\bm{x_0})$ is white noise added to the data.  The noise is added by specifying a $\mathrm{SNR} =$ $10$ as described in equation 6.  The data $\bm{y_0}$ is shown in Figure 4.

\begin{figure}
\centering
  \includegraphics[width=0.45\textwidth]{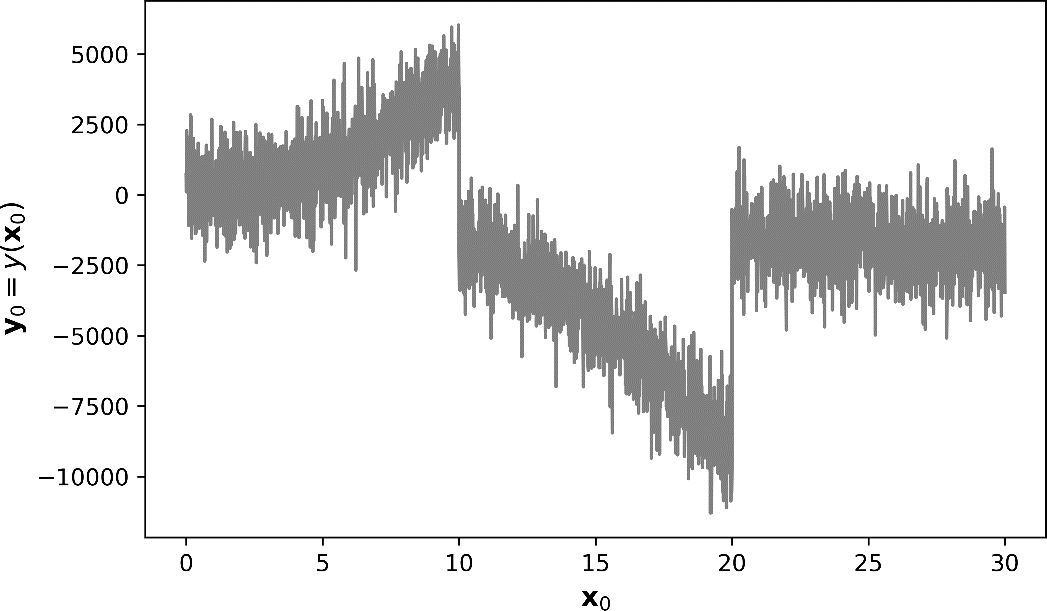}
  \caption{Example dataset for 3-domain system. There are two boundaries that must be discovered for the algorithm to find the simplest model that accurately describes the entire dataset.}
  \label{fig:subvol}
\end{figure}

As in the previous example, a power series model is employed with $K \leq 2$ where larger exponents correspond to greater complexity.  During the first step of the search process, the algorithm searches for a boundary by trying each point in $\bm{x_0}$ in the original domain $D_0$.  Figure 5 shows the first search step.

\begin{figure}
\centering
  \includegraphics[width=0.45\textwidth]{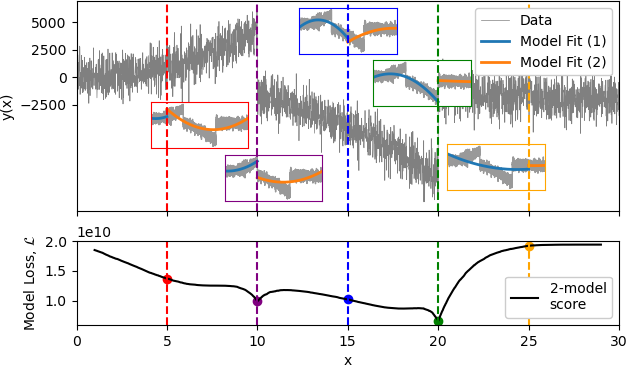}
  \caption{The first search step to find a boundary in a multi-domain data set.  The minimum 2-model loss, shown by the dashed green vertical line and green point in the loss line, corresponds to a boundary found in the data.}
  \label{fig:subvol}
\end{figure}

The algorithm identifies the point $x=20$ as a boundary hyperplane, splitting the domain $D_0$ into $x \in [0,20)$ for subdomain $D_1$  and $x \in [20,30]$ for subdomain $D_2$.  Once the original domain $D_0$ has been separated into $D_1$ and $D_2$, the search process continues by repeating the algorithm on the two subdomains $D_1$ and $D_2$ as shown in Figure 1.  The search process fails to find any additional subdomains within $D_2$.  However, there is an additional boundary discovered in the domain $D_1$.  The search in domain $D_1$ for the boundary at $x=10$ is shown in Figure 6.  The algorithm correctly identifies the boundary at $x=10$ and consequently divides $D_1$ into $D_3$ and $D_4$.  The algorithm searches $D_3$ and $D_4$ but fails to find additional boundaries. 

\begin{figure}
\centering
  \includegraphics[width=0.45\textwidth]{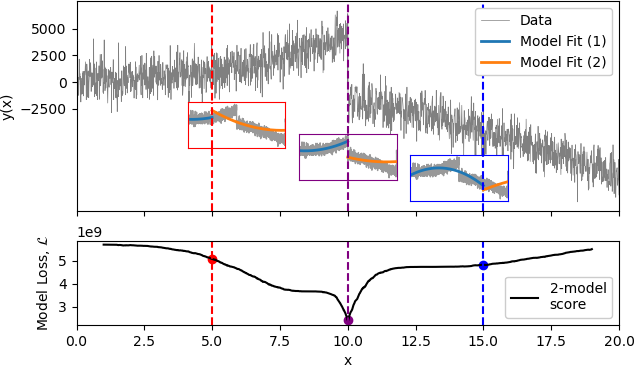}
  \caption{Second search step of the data shown in Figure 5.  The boundary $x=10$ is identified as an optimum hyperplane separation, leading to discovery of domains $D_3$  and $D_4$.}
  \label{fig:subvol}
\end{figure}

\subsection{Domain Discovery in Stress-Strain Data}
This section shows the 1-dimensional RDP algorithm applied to noisy stress-strain data obtained by deforming amorphous carbon in a molecular dynamics simulation.  The stepwise strain values create a challenging dataset for stress-strain domain identification.  The goal is to identify the physical regimes of the deformation curve: elastic deformation, plastic deformation, and failure.  Details on the simulation are provided in the Appendix.

This example uses the same set of power series models from equation 8.  The penalty function is \begin{equation}
p(C(f(x,K))=1-0.1(2-K).
\end{equation}Figure 7 shows the regions found by the algorithm in several datasets.  Example (a) shows a clear separation into the expected regions of a stress-strain curve: elastic (linear) region, plastic deformation (the curved region at the top), and failure (what happens after the maximum is reached).  Examples (b), (c), and (d) transition from elastic mode into failure mode without a large plastic region.  The failure modes are often decomposed into additional regions as shown in example (b).  The overall results are comparable to the regions that might be drawn by a human observer.

\begin{figure}
\centering
  \includegraphics[width=0.45\textwidth]{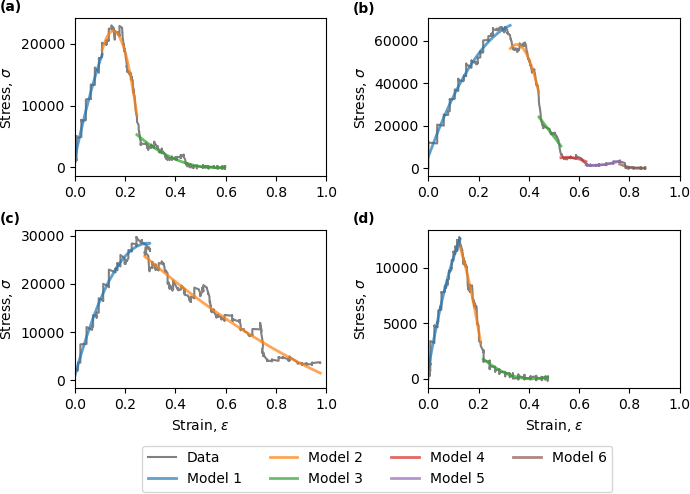}
  \caption{The domain discovery procedure works on noisy data, including tensile test stress-strain data generated by molecular simulations.  The data in this figure was prepared using simulations of amorphous carbon deformation performed with LAMMPS.  Despite the varied characteristics of examples (a) through (d), the algorithm is able to identify the physical regimes corresponding to elastic deformation and failure.}
  \label{fig:subvol}
\end{figure}

\subsection{Implementation:  2D Example}
The RDP method is demonstrated in 2D by first showing a piecewise 2D scalar function, then applying the procedure to a vectorized two-dimensional fluid dynamics data set.  This section illustrates how to implement the algorithm in higher dimensions, including how to enumerate possible hyperplane boundaries and training models.

Consider the model
\begin{align}
&z(x,y) = \left\{\begin{matrix}
z_1(x,y), & \:\:\:\: x+y \leq 1.05 \\
z_2(x,y), & \:\:\:\: x+y > 1.05 \\
\end{matrix}\right. \nonumber \\
&z_1(x,y) = -15x^2+3x+4y^2 \\
&z_2 (x,y) =  -10x+12y^3. \nonumber
\end{align}
The line $y=-x+1.05$ separates the data based on the coordinates $x$ and $y$.  The data $z(x,y)$ and the functions $z_1 (x,y)$ and $z_2 (x,y)$ are shown in Figure 8.

This example employs a multi-dimensional power series model \begin{equation}
f(x,y,K,J) = \sum_{k=0}^{K} \sum_{j=0}^{J} a_{j,k} x^k y^j
\end{equation} where $a_{j,k}$ is a learned coefficient for the term $x^k y^j$.  The power series contains $J \cdot K$ total terms.  Unlike the 1D case, only one model is considered for these 2D systems.  Similar to the 1D case, the algorithm executes a brute force search on all linear separations of the data. 

A partition hyperplane of dimension $N-1$ for a 2-dimensional dataset is a line.  Therefore, it is necessary to identify all the unique separations of the 2D data by a boundary line.  If the 2D data is on a rectangular grid, the unique boundary lines can be enumerated by finding the set of all lines that intersect two points along the grid perimeter.

Figure 9 shows the result of applying all possible linear separations to the data using lines defined by sets of two intersection points along the domain perimeter.  The algorithm discovers the lowest error corresponds to a pair of points defining a line that intersects edge 3 and edge 4, which is the correct line indicating the true separation of the data $(y=-x+1.05$.  Details related to the power series regression and separation of the data are provided in the Appendix.

\subsection{Balance Discovery in 2D Vectorized Flow-Field Data}
The RDP method is demonstrated on physically realistic vectorized data using a fluid dynamics example.  This problem comes from the study of boundary layer theory, which recognizes a distinct difference in the flow dynamics of a fluid near a physical boundary and away from a physical boundary.  The system is described by the Reynolds-averaged Navier-Stokes (RANS) equation
\begin{equation}
\bar{u} \frac{\partial \bar{u}}{\partial x} + \bar{v} \frac{\partial \bar{v}}{\partial y} = \rho^{-1} \frac{\partial \bar{p}}{\partial x} + v \triangledown^2 \bar{u} - \frac{\partial}{\partial y} \bar{u}' \bar{v}' - \frac{\partial}{\partial x} \overline{u'^2}
\end{equation} where $u$ and $v$ are components of the velocity vector $\bm{u}=(u,v,w)$, $p$ is pressure, $\rho$ is fluid viscosity, and $x$ and $y$ are spatial coordinates.  RANS is a time-averaged equation for fluid flow resulting from the Reynolds decomposition, which breaks the observables (velocity and pressure) into time-averaged and fluctuating components.  The RANS formulation in this example is focused on the time-averaged rather than fluctuating components.  The goal of the RDP procedure is to identify a linear separation in the data which identifies significantly different behaviors in fluid flow in different regions.  We use data from a numerical simulation of the boundary layer flow dynamics [23].

\begin{figure}
\centering
  \includegraphics[width=0.45\textwidth]{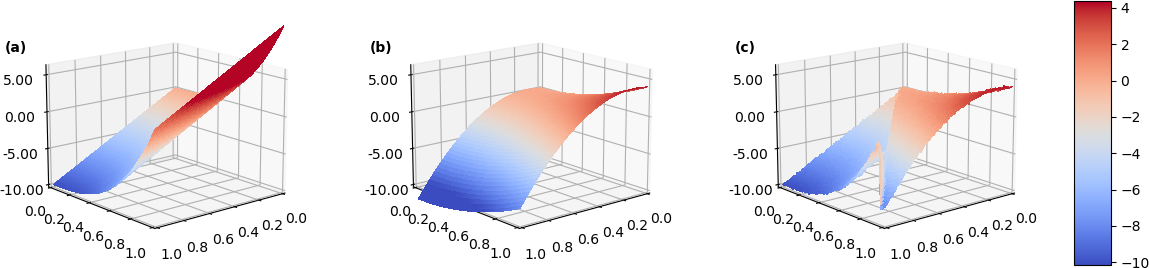}
  \caption{Panel (a) shows the function $z_1(x,y)$, panel (b) shows $z_2(x,y)$, and panel (c) shows $z(x,y)$, which is a piecewise function of $z_1$ and $z_2$ separated by the line $y=-x+1.05$.}
  \label{fig:subvol}
\end{figure}

\begin{figure}
\centering
  \includegraphics[width=0.45\textwidth]{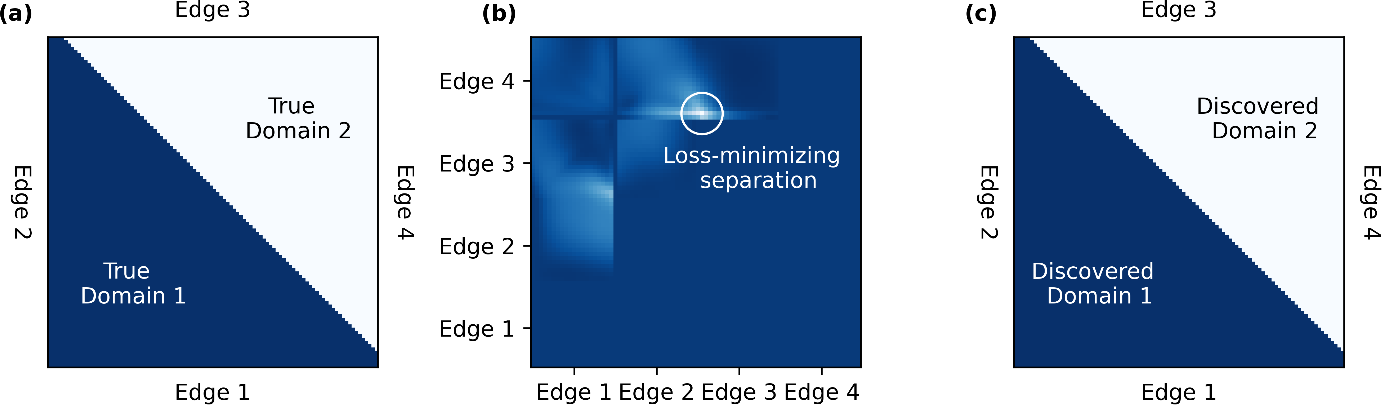}
  \caption{True linear separation in data (a), the discovered separation in the data (c), and the loss surface for fitting models defined b the separation in data (b).  Each pixel in (b) corresponds to the 2-model loss for a line connecting two perimeter points.  The location of points corresponding to each edge is identified for clarity.  The error-minimizing 2-domain model is described by a boundary line separating the domains that connects edge 3 and edge 4.}
  \label{fig:subvol}
\end{figure}

The power series model shown in equation 15 is now formulated with an output variable for each of the vector components of the fluid flow.  The model is fit to each domain using multivariate regression.  Additional details are provided in the Appendix.

\begin{figure}
\centering
  \includegraphics[width=0.45\textwidth]{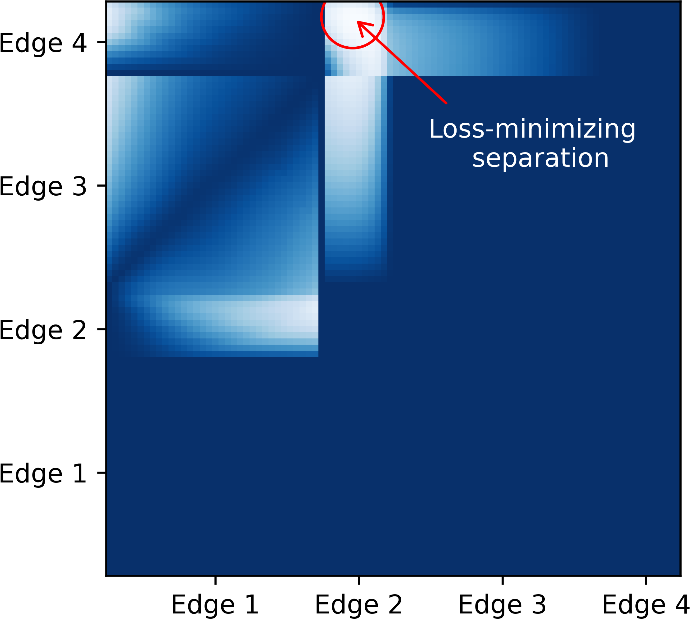}
  \caption{Loss surface for the fluid dynamics model.  The optimal linear separation in the data occurs as a connection between edges 2 and 4, as shown in Figure 10.}
  \label{fig:subvol}
\end{figure}

The loss surface for this example is shown in Figure 10.  The optimal linear separation identified by the RPD algorithm intersects edges 2 and 4 (the left and right sides, respectively) of the data set.  Note that many possible linear separations of the dataset satisfy the criterion in equation 4, but this separation is found to be optimal.  Interestingly, there are some other boundaries with relatively low error that connect edges 1-2, 1-3, and 2-3.  These boundaries typically define a domain near the $x=0$ region, which is unsurprising as this is a region of relatively low pressure in this model.  However, because the model focused on learning relationships for the velocity vector, these are sub-optimal models when compared to the division which separates a constant velocity free-flow region and the region of complex flow.

Figure 11 shows the optimum linear separation of the flow vector field into two domains.  The learned boundary separates a region of nearly constant flow velocity, shown in yellow, from a region with more complex flow dynamics, shown in green and blue. 
\section{Conclusion}

This work establishes a Recursive Domain Partitioning (RDP) method for identifying subdomains of data using model precision as a metric. The subdomains are defined by linear separations of data using hyperplanes. Each hyperplane division is discovered sequentially in a binary space partitioning procedure to uncover the optimal subdomain definitions. The method can discover minimally-complex models in order to avoid overfitting, prevent increasing complexity without increasing accuracy, and use complex relationships only where needed. We show the method on 1D and 2D data sets, including vectorized 2D data. 

A number of improvements may be pursued in the future. The concept was proven here using power series models, although polynomials or any type of model is compatible with the principal method. For example, future studies could utilize neural networks of varying complexity. Additionally, the RDP approach is currently one-directional and can only increase the number of domains.  Future development on this method should include the ability to decrease the number of subdomains by combining subdomains with similar models.

\begin{figure}
\centering
  \includegraphics[width=0.45\textwidth]{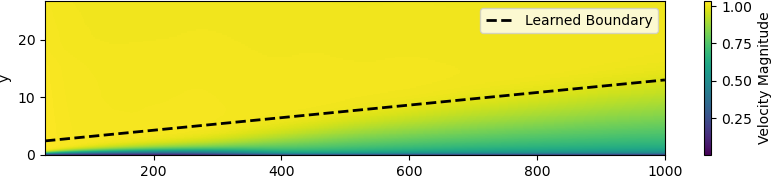}
  \caption{Discovered boundary in the fluid dynamics flow model.  The plot shows the magnitude of the velocity vector u.  The learned boundary separates the free-flow region, consisting of nearly constant flow velocity, from the region where more complex flow dynamics occur.}
  \label{fig:subvol}
\end{figure}

The largest drawback to the currently proposed method is the brute-force nature of the search algorithm. Currently, the algorithm searches every possible hyperplane boundary and in order to identify the best boundary.  However, it may be possible to implement a nonlinear optimization method to reduce the computational resources required to enumerate and test all possible hyperplanes.  These may include coarse- to fine-grained grid searches, evolutionary algorithms, or simulated annealing. It may also be interesting to combine dimensionality reduction techniques or manifold learning approaches with the RDP procedure to reduce computational complexity of the brute-force approach.

\section{Appendix}
\subsection{Stress-strain tensile deformation simulation in LAMMPS}

The stress-strain data is derived from LAMMPS simulations of amorphous carbon structures.  The initial states of the LAMMPS simulations are determined by randomized carbon atom placements inside of a test volume. The volume is fixed at one wall and pulled apart in tensile elongation along a single axis.  Quantifying stress-strain in an amorphous structure is challenging because (a) the simulation cannot be configured to fairly assume periodic boundary conditions, and (b) fixing some atoms at one edge of the box skews the strain values. 

\subsection{Boundary hyperplanes in 2D}
The data points in 2D are located on a rectangular grid with equal spacing on both axes.  An example is shown in Figure 12.  This type of grid is common for studying differential equations and is straightforward to implement experimentally.

Every line that can divide this dataset must pass through two edges of the measurement grid.  We can systematically identify all possible boundary lines using the discrete points along the edges of the measurement grid.  Figure 13 shows a sample of partition lines connecting points on the edges.

\subsection{Data Selection in 2D}
Each prospective boundary line divides the domain into two subdomains.  Colloquially, one of the subdomains is "on one side" of the line and the other subdomain is "on the other side".  The orientation test from computational geometry is used to make this distinction.  Given a line intersecting two points $a$ and $b$, and a third point $c$ off the line: \begin{equation}
R(a,b,c)= \mathrm{sign} \left ( \left | \begin{matrix}
a_x & a_y & 1 \\ 
b_x & b_y & 1 \\ 
c_x & c_y & 1
\end{matrix} \right | \right )
\end{equation} where $a_x$ and $a_y$ are the $x$ coordinate and $y$ coordinate of point $a$, respectively, assuming a Cartesian coordinate system.  The sign of the determinant indicates the location of point $c$ is relative to the line $\overline{ab}$.  All points on one side of the line produce a positive determinant, while all points on the other side produce a negative determinant.  The point falls on the line if the determinant is zero.  Using this test to develop an indicator function allows a program to independently access data on either side of the boundary line $\overline{ab}$.

\begin{figure}
\centering
  \includegraphics[width=0.45\textwidth]{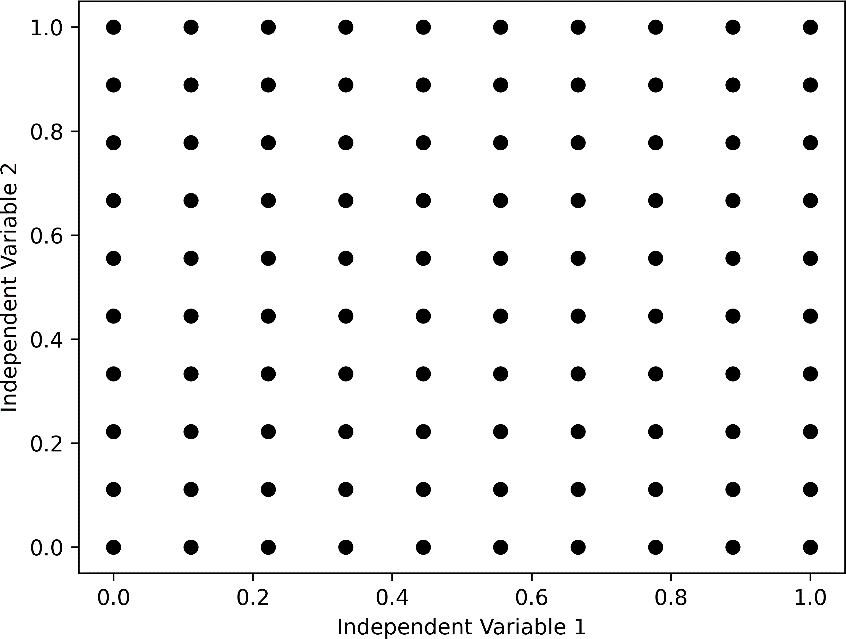}
  \caption{Example measurement grid in 2D.  The black points indicate measurement points for the 2D data.  The variables are discretized in the range from 0 to 1 with step size of $0.1$.}
  \label{fig:subvol}
\end{figure}

\subsection{Power Series Symbolic Regression in 2D}

The multivariate power series used for 2D data analysis is slightly more complicated than the 1D case.  Polynomial fitting using least-squares regression in 1D is relatively well-known. The multivariate power series model used in 2D is \begin{equation}
f(x,y) = \sum_{k=0}^{3} \sum_{j=0}^{3} a_{j,k} x^k y^j.
\end{equation}Models up to cubic order with respect to x and y are used.  This indicates a multivariate power series with 16 total terms: $1, x, x^2, x^3, y, y^2, y^3, xy, xy^2, xy^3, x^2 y, x^2 y^2, x^2 y^3, x^3 y, \\x^3 y^2, x^3 y^3$.  The least-squares regression method uses a feature column for each of the terms in the power series.  For a set of measurements $z(x,y)$ the equation is:
\begin{center}
$\begin{bmatrix}
z(x_1,y_1)\\ 
\vdots \\ 
z(x_n,y_1)\\ 
\vdots \\ 
z(x_n,y_m)
\end{bmatrix}
=$
\end{center}
\begin{equation}
\begin{bmatrix}
1 & x_1 & \cdots & y_1 & \cdots & x_1 y_1 & \cdots & x_1^3 y_1^3\\ 
\vdots & \vdots & \ddots & \vdots & \ddots & \vdots & \ddots & \vdots \\ 
1 & x_n & \cdots & y_1 & \cdots & x_n y_1 & \cdots & x_n^3 y_1^3\\ 
\vdots & \vdots & \ddots & \vdots & \ddots & \vdots & \ddots & \vdots \\ 
1 & x_n & \cdots & y_m & \cdots & x_n y_m & \cdots & x_n^3 y_m^3
\end{bmatrix}
\bm{\Xi}
\end{equation}
where the coefficient vector $\bm{\Xi}$ is learned by least squares regression.

For the fluid flow boundary layer dataset, the measurements are comprised of the vector $\bm{u}=\left ( u(x,y),v(x,y) \right )$.  This example employs the following vectorized multivariate regression matrix:
\begin{center}
$\begin{bmatrix}
u(x_1,y_1) & v(x_1,y_1) \\ 
\vdots & \vdots \\ 
u(x_n,y_1) & v(x_n,y_1) \\ 
\vdots & \vdots \\ 
u(x_n,y_m) & v(x_n,y_m)
\end{bmatrix}
=$\end{center} \begin{equation}
\begin{bmatrix}
1 & x_1 & \cdots & y_1 & \cdots & x_1 y_1 & \cdots & x_1^3 y_1^3\\ 
\vdots & \vdots & \ddots & \vdots & \ddots & \vdots & \ddots & \vdots \\ 
1 & x_n & \cdots & y_1 & \cdots & x_n y_1 & \cdots & x_n^3 y_1^3\\ 
\vdots & \vdots & \ddots & \vdots & \ddots & \vdots & \ddots & \vdots \\ 
1 & x_n & \cdots & y_m & \cdots & x_n y_m & \cdots & x_n^3 y_m^3
\end{bmatrix}
\bm{\Xi}.
\end{equation}

\begin{figure}
\centering
  \includegraphics[width=0.45\textwidth]{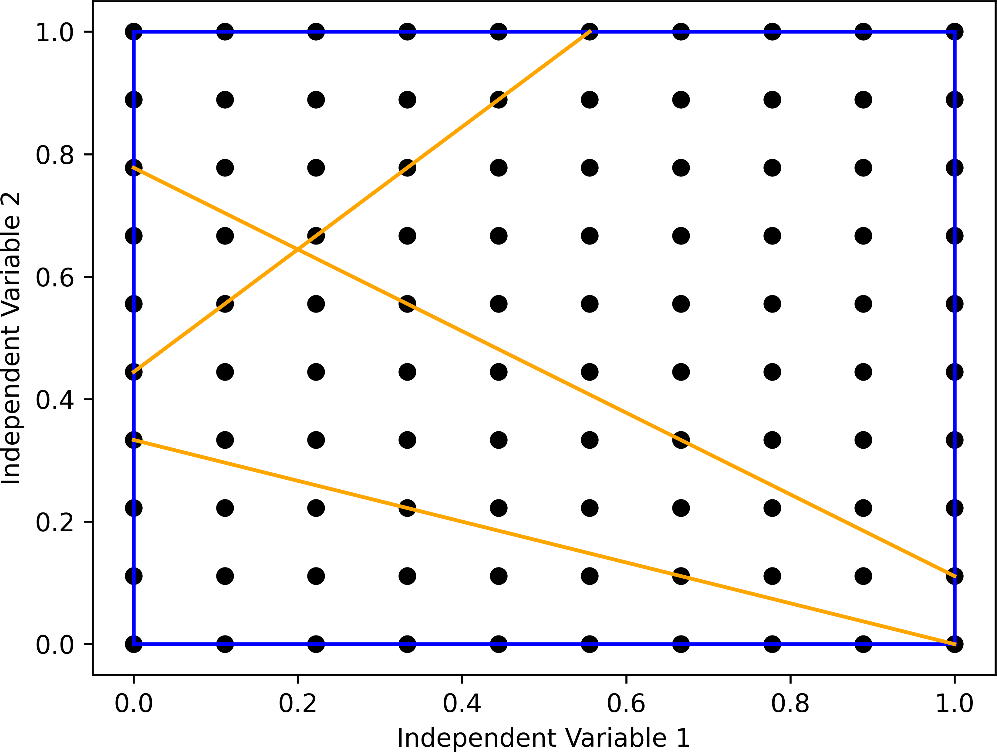}
  \caption{2D Measurement grid with highlighted boundary edges (blue) and example boundary lines (orange).  All possible boundary lines can be enumerated by identifying each unique combination of two points along the edges and drawing lines which intersect those points.}
  \label{fig:subvol}
\end{figure}

\nocite{*}
\bibliographystyle{unsrt}
\bibliography{biblio}

\end{document}